\documentclass[twocolumn,aps,preprintnumbers,prl,showpacs,
amsfonts,amsmath,amssymb,floatfix]{revtex4}

\newtheorem{thm}{Theorem}{}{}
\newcommand{\diffD}{{\mathrm d}}
\def\ket#1{\mathinner{|{#1}\rangle}}

\begin{document}

\title{Black hole evaporation rates without spacetime}

\author{Samuel L.\ Braunstein}
\affiliation{Computer Science, University of York, York YO10 5DD, United Kingdom}
\author{Manas K.\ Patra}
\affiliation{Computer Science, University of York, York YO10 5DD, United Kingdom}

\date{\today}

\begin{abstract}
Verlinde recently suggested that gravity, inertia, and even spacetime
may be emergent properties of an underlying thermodynamic theory. This
vision was motivated in part by Jacobson's 1995 surprise result that
the Einstein equations of gravity follow from the thermodynamic properties
of event horizons. Taking a first tentative step in such a program, we
derive the evaporation rate (or radiation spectrum) from black hole event
horizons in a spacetime-free manner. Our result relies on a Hilbert space
description of black hole evaporation, symmetries therein which follow
from the inherent high dimensionality of black holes, global conservation
of the no-hair quantities, and the existence of Penrose processes. Our
analysis is not wedded to standard general relativity and so should
apply to extended gravity theories where we find that the black hole
area must be replaced by some other property in any generalized area
theorem.
\end{abstract}


\maketitle

Jacobson's derivation of the Einstein field equations \cite{Jacobson,B}
indicates that the physics across event horizons determines the 
structure of gravity theory. To realize Verlinde's \cite{Verlinde} vision,
therefore, it seems reasonable to start by studying this physics, at the 
microscopic or quantum mechanical level, i.e., the particle
production by the event horizon \cite{Jacobson,B,Verlinde}. Although we 
focus primarily on black hole event horizons, much of the analysis should 
apply to general event horizons.

We require that the particle production mechanism be consistent with 
the complete unitary evaporation of a black hole.  
This strongly suggests tunneling as this mechanism
\cite{NikolicA,BraunsteinA,supmat}: Particles quantum mechanically tunnel
out across the classically forbidden barrier associated with the event
horizon to emerge as Hawking radiation \cite{Kraus,ParikhWilczek}.
Recent calculations of tunneling probabilities (the evaporation rate)
from black holes have incorporated backreaction from an
escaping particle on the classical black hole due to conservation laws
of the no-hair quantities. Although only a limited number of black hole
types, particle types, and WKB trajectories have been studied in this
way (see e.g., \cite{Kraus,ParikhWilczek,a}) the tunneling probabilities
appear to take the generic form
\begin{equation}
\label{eq:formTunProb}
\Gamma \propto e^{{\cal S}_{\text{final}}-{\cal S}_{\text{initial}}},
\end{equation}
where ${\cal S}_{\text{initial(final)}}$ are the thermodynamic entropies
of the black hole before (after) the tunneling process. 
In the simplest
case of a spinless particle of energy $\varepsilon$ evaporating from a
Schwarzschild black hole of mass $M$ along a radial trajectory, one has
\cite{ParikhWilczek}
\begin{equation}
\label{eq:STunProb}
\Gamma \propto e^{4\pi(M-\varepsilon)^2-4\pi M^2}.
\end{equation}

Tunneling moves subsystems from the black hole interior (int) to the
exterior, appearing as radiation (R) \cite{BraunsteinA}. Formally, the
simplest Hilbert space description of such a process is given by
\begin{equation}
|i\rangle_{\text{int}} \rightarrow (U|i\rangle)_{BR}.
\label{evapA}
\end{equation}
Here $B$ denotes the reduced size subsystem corresponding to the
remaining interior of the black hole, $|i\rangle$ is the initial state of
the black hole interior (which we take here to be pure for convenience
and without loss of generality \cite{BraunsteinA}), and $U$ denotes the
unitary process ``selecting'' the subsystem to eject.

Note that spacetime and black hole geometry are not explicit in
Eq.~(\ref{evapA}). Even the event horizon appears only as a generic
Hilbert space tensor product structure separating what we call interior
from exterior \cite{Hawking76a,supmat}. These observations provide support
for the conjecture that Eq.~(\ref{evapA}) should apply to evaporation
across {\it arbitrary\/} event horizons. For a more detailed motivation
and history of this description, see Refs.~\cite{BraunsteinA,supmat}.

We show that Eq.~(\ref{evapA}), symmetries therein, and global conservation
laws imply Eq.~(\ref{eq:formTunProb}) for evaporation across black hole
event horizons. To apply the generic Eq.~(\ref{evapA}) specifically to
such horizons we assume a correspondence between quantum and classical
descriptions of black holes. In particular, we rely on their labeling
by no-hair quantities and the existence of Penrose processes.

The first symmetry we investigate is a permutation symmetry in the order of
``decay'' products (evaporated particles). Consider a pair of distinct
subsystems of the radiation. Interchanging them corresponds to a unitary
operation which may be formally absorbed into the internal unitary $U$
in Eq.~(\ref{evapA}). Because the Hilbert space dimensionalities needed
to describe a black hole are so vast (at least ${10^{10}}^{77}$ for
a stellar-mass black hole) random matrix theory \cite{Levy,BraunsteinA}
tells us that the {\it statistical\/} behavior of Eq.~(\ref{evapA}) is
excellently approximated by treating $U$ as a random unitary (i.e.,
by using a Haar average). Therefore, permuting the order in which
particles appear as Hawking radiation will have no statistical effect.

In the case of a Schwarzschild black hole of mass $M$ undergoing a pair
of consecutive evaporation events producing spinless particles in an
$s$ wave of energies $\varepsilon_1$ and $\varepsilon_2$, this permutation
symmetry implies an equality between tunneling probabilities,
$\Gamma(\varepsilon_1,\varepsilon_2|M)
\!=\!\Gamma(\varepsilon_2,\varepsilon_1|M)$,
or in terms of conditional probabilities
\begin{equation}
\Gamma(\varepsilon_1|M)\, \Gamma(\varepsilon_2|\varepsilon_1,M)
=\Gamma(\varepsilon_2|M)\, \Gamma(\varepsilon_1|\varepsilon_2,M).
\end{equation}

Now in field theory calculations, tunneling probabilities concern
transitions between classical macroscopic spacetime geometries.
Abstracting this into a spacetime free language we would say that
earlier decays should only affect subsequent decays through their
backreaction on the black hole's identity via conservation laws. In
the simple scenario above, particles only carry away black hole mass
as energy, so
$\Gamma(\varepsilon|\varepsilon',M)=\Gamma(\varepsilon|M-\varepsilon')$.
This leaves the single-particle functional relation
\begin{equation}
\label{eq:ampScalarFunc1}
\Gamma(\varepsilon_1|M)\, \Gamma(\varepsilon_2|M-\varepsilon_1)
=\Gamma(\varepsilon_2|M)\, \Gamma(\varepsilon_1|M-\varepsilon_2).
\end{equation}

\begin{thm}
\label{thm1}
Suppose the function $\Gamma$ is continuously differentiable in
its domain of definition and satisfies Eq.~\eqref{eq:ampScalarFunc1}.
Then its general solution is
\begin{equation}
\Gamma(\varepsilon|M)= e^{f(M-\varepsilon)-f(M)+h(\varepsilon)},
\label{eq7}
\end{equation}
where $f$ and $h$ are arbitrary functions, continuously differentiable
except possibly at some boundary points.  
\end{thm}

The general solution provided by this functional equation (see
Ref.~\cite{supmat} for proofs of all our theorems) easily matches
the known result of Eq.~(\ref{eq:STunProb}). Thus at least for this
scenario, the permutation symmetry predicted by the Hilbert space
description of Eq.~(\ref{evapA}) is supported by quantum field theoretic
tunneling calculations on curved spacetime \cite{ParikhWilczek}.
(Consistency with Hawking's original result of a thermal distribution
for black hole radiation \cite{Hawking75} when backreaction
is negligible, i.e., when the energy $\varepsilon$ carried away is
infinitesimal, would immediately implicate $f$ as the black hole's
thermodynamic entropy.)

To generalize this result to more general scenarios we need only
assume that any changes that occur in an event horizon's identity due
to evaporation are characterized by {\it linear\/} conservation laws.
We will now explicitly show that this approach is valid for evaporation
of black holes.

Recall that the no-hair theorem \cite{nohair} tells us that a
black hole is characterized solely by its mass $M$, charge $Q$, and
angular momentum $J$ along some axis $\hat n$. The parameters
$\vec X\equiv (M,Q,J)$ can be ``readout'' \cite{Bek} (copied) by
arbitrarily many observers throughout the
spacetime geometry --- they correspond to {\it classical information\/}
about the quantum state of the black hole. It is therefore natural
to associate a classical black hole with a quantum state which is
the simultaneous eigenstate of $M$, $Q$, and $J_{\hat n}$. To ensure
that angular momentum is described by a single quantum number (as required
by the no-hair theorem), the angular momentum state for a black hole
must correspond to a spin-coherent state
$|J,J\rangle_{\hat n} = R(\theta,\phi)
|j\!=\!J,m\!=\!J\rangle$,
where $|j,m\rangle$ are the usual simultaneous eigenstates of total
angular momentum $J^2$ and $J_{\hat z}$ and $R(\theta,\phi)$ is a rotation
operator which maps $\hat z$ to $\hat n$. This correspondence has the
added feature of making the quantum description of a black hole a
minimum uncertainty angular momentum state --- i.e., as classical
as possible in its angular momentum degrees of freedom.

Now the ability of an infinite set of observers throughout spacetime
to copy the classical information about the (black hole) geometry 
they are sitting in places a very strong constraint on any physical
process. In particular, any process that yields a superposition of
black hole states can only preserve this ``copyability'' if the
superposition can be expressed as a sum over mutually orthogonal black
hole states. This property and the presumed conservation during 
black hole evaporation of total energy, charge, and angular momentum yield
the following. 

\begin{thm}
\label{thm2}
Consider a lone black hole $\vec X\equiv (M,Q,J)$ oriented along some
direction $\hat n$ that undergoes an evaporative process
yielding a particle and a daughter black hole. If the particle's
energy $\varepsilon$, charge $q$, and total (spatial plus spin) angular
momentum $j$ along the $\hat n$ axis are measured, then the remaining
state of the daughter black hole will be described by the no-hair
triple $\vec X - \vec x$ along $\hat n$, where
$\vec x\equiv (\varepsilon, q, j)$. (Note, $\hat n$ is arbitrary
for $J=0$.)
%
\end{thm}

This theorem tells us that the transition from black hole mother to
daughter by evaporation satisfies a simple set of linear conservation
laws. Note, Theorem~\ref{thm2} should not be taken to imply that the
particle is fully described by $\vec x\equiv (\varepsilon, q, j)$ and
has no other ``hair.'' For example, the particle need not be in an overall
spin-coherent state. 

It immediately follows from Theorem~\ref{thm2} and the permutation
symmetry already discussed that the probability
$\Gamma(\vec x|\vec X)$ for a particle with triple
$\vec x\equiv (\varepsilon, q, j)$ to tunnel from a black hole
with no-hair triple $\vec X=(M,Q,J)$ will satisfy
$\Gamma(\vec x|\vec X)\,\Gamma(\vec x'|\vec X - \vec x)
=\Gamma(\vec x'|\vec X)\,\Gamma(\vec x|\vec X - \vec x')$,
again presuming that prior evaporative events only affect subsequent
decays via their (linear) conservation laws, so
$\Gamma(\vec x|\vec x',\vec X)=\Gamma(\vec x|\vec X-\vec x')$. It is
therefore natural to extend Theorem~\ref{thm1} to the multivariate case.

\begin{thm}
\label{thm3}
Let $\Gamma(\vec x|\vec X)$ be a positive real function that is
continuously differentiable in each of the variables,
$\vec x, \vec X\in D \subset \mathbb{R}^n$, where the domain $D$ of
each of the (vector) arguments is assumed be to a closed subset of
$\mathbb{R}^n$. Suppose that $\Gamma$ satisfies the functional equation 
\begin{equation}
\label{eq:basicFunc1_multi}
\Gamma(\vec x|\vec X)\,\Gamma(\vec x'|\vec X - \vec x)
=\Gamma(\vec x'|\vec X)\,\Gamma(\vec x|\vec X - \vec x').
\end{equation}
Then the function $\Gamma$ is given by 
\begin{equation}
\label{tunProb}
\Gamma(\vec x|\vec X)= e^{f(\vec X-\vec x)-f(\vec X)+h(\vec x)},
\end{equation}
where $f(\vec X)$ and $h(\vec x)$ are continuously differentiable
functions of their arguments.
\end{thm}

Moving beyond black holes, if a general event horizon were characterized
by some vector of attributes $\vec X$ and if evaporation
produced a linear backreaction to these attributes, then permutation
symmetry and this theorem would imply the same functional form quite
generally. Again, were we to assume consistency with a thermal spectrum
\cite{Hawking75} when backreaction is negligible, we would find that
$f(\vec X)$ is the thermodynamic entropy associated with the event
horizon. Instead of this route, let us consider another symmetry
specifically possessed by black holes that will allow us to determine
both $f$ and $h$ and to uncover further structure. This will have the
added benefit of allowing us to infer an almost certain breakdown of
the area theorem \cite{AreaThm} in extended gravity theories.

In quantum theory a reversible process should be represented
by a unitary operation in Hilbert space. A reversible Penrose
process \cite{Pen1} allows one to freely interconvert between
two black holes with no-hair triples $\vec X_1$ and $\vec X_2$, provided
only that their irreducible masses ${\cal I}$ are equal.
Consider a pair of such reversible processes applied to a black hole so
as to bracket a single tunneling event. If the unitary describing that
event can be well approximated by its Haar average we must have
\begin{equation}
\Theta(\vec X_1,\vec X_1') =\Theta(\vec X_2,\vec X_2'),
\end{equation}
whenever ${\cal I}(\vec X_1)={\cal I}(\vec X_2)$ and
${\cal I}(\vec X_1')={\cal I}(\vec X_2')$. Where for convenience we
introduce {\it transition probabilities\/}
$\Theta(\vec X,\vec X')\equiv \Gamma(\vec X - \vec X' | \vec X)$
for a mother black hole $\vec X$ to yield a daughter $\vec X'$ after
a single tunneling event.

A holographic view of an event horizon might be stated as saying that
the Hilbert space beyond the event horizon is effectively encoded
entirely at or near the event horizon itself. In such a case,
the Haar symmetry for {\it each\/} individual tunneling event
would be an ideal description of the random sampling of near
event-horizon degrees of freedom for ejection as radiation. Specifically
for black holes, Ref.~\cite{BraunsteinA} proves that, in order
to preserve the equivalence principle during (unitary) evaporation over
a black hole's lifetime, virtually the entire Hilbert space of the black hole
must be encoded at the surface (in the form of trans-event-horizon
entanglement), with at most a vanishingly small proportion located
within the black hole interior. In this case Haar symmetry for {\it each\/}
tunneling event would still be an excellent approximation. (In contrast,
if the Hilbert space of the black hole interior were not encoded
primarily near the surface, then its Haar invariance would require the
dynamical assumption of a very short ``global thermalization time'' for
the black hole --- little more than the time for charge to spread across
a black hole's surface.)

\begin{thm}
\label{thm4}
Let ${\cal I}:\Sigma\rightarrow \mathbb{R}$ be continuous on $\Sigma$
(a closed subset of $\mathbb{R}^n$) and continuously differentiable
on its (nonempty) interior $\Sigma^o$. Assume further that the subset
$K\subset\Sigma^o$ on which all partial derivatives
$\partial {\cal I}/\partial X_i$ vanish contains no open set.
Furthermore, let $\Theta:\Sigma\times \Sigma\rightarrow \mathbb{R}$
be continuous. Suppose 
\begin{eqnarray}
&&{\cal I}(\vec X_1)={\cal I}(\vec X_2) ~~\text{and}~~
{\cal I}(\vec X_1')={\cal I}(\vec X_2')\nonumber\\
&&~~\implies \Theta(\vec X_1,\vec X_1') =\Theta(\vec X_2,\vec X_2'),
\label{hypot}
\end{eqnarray}
then $\Theta(\vec X,\vec X') =\theta({\cal I}(\vec X),{\cal I}(\vec X'))$
for some function $\theta$.
\end{thm}

Combining this with Theorem~\ref{thm3} implies that the universal
function $f(\vec X)=u({\cal I}(\vec X))$ is some function solely of
the irreducible mass ${\cal I}$ and that $h(\vec x)$ must be a constant. In
other words,
\begin{equation}
\Gamma(\vec x|\vec X) ={\cal N}
e^{u({\cal I}(\vec X-\vec x))-u({\cal I}(\vec X))},
\label{simpleform}
\end{equation}
where ${\cal N}$ is a normalization constant. 

We will now determine the function $u$ from the fact that the Hilbert
space description of evaporation in Eq.~(\ref{evapA}) is manifestly
reversible. We assume that a black hole can evaporate away completely,
since any stable black hole remnant would itself be tantamount to
a failure of unitarity \cite{Giddings} and hence of quantum mechanics.
Consider therefore the {\it complete\/} evaporation of our Hilbert
space black hole with initial no-hair triple $\vec X$, leaving nothing
but radiation. The probability for seeing a specific stream of radiation
with triples $\{\vec x_1, \vec x_2,\ldots\}$ may be precisely computed from
Eq.~(\ref{simpleform}) as
\begin{equation}
\Gamma(\vec x_1, \vec x_2,\ldots|\vec X)
\equiv {\cal N}' e^{u({\cal I}(\vec 0))-u({\cal I}(\vec X))},
\label{radiation}
\end{equation}
where $u({\cal I}(\vec 0))$ must be finite to ensure that complete
evaporation is possible. Because Eq.~(\ref{radiation}) is independent
of the specific radiation triples, it implies that the
{\it thermodynamic\/} entropy of the radiation is exactly 
\begin{equation}
{\cal S}_{\text{rad}}=u({\cal I}(\vec X))-u({\cal I}(\vec 0))-\ln {\cal N}',
\label{eq18}
\end{equation}
(see Refs.~\cite{ZhangA,SingletonA} for related arguments).
To ensure reversibility the entropy in the radiation must equal
the thermodynamic entropy of the original black hole ${\cal S}(\vec X)$.
In other words, we must have ${\cal S}_{\text{rad}}={\cal S}(\vec X)$,
which in turn implies that $u({\cal I}(\vec X))$ is just the
thermodynamic entropy associated with the black hole.

In general relativity both the thermodynamic entropy of a black hole and
its irreducible mass are known to be functions of the black hole surface
area, so their connection here is not too surprising. However, in
higher curvature theories of gravity, Gauss-Bonnet gravity and other
Lovelock extended gravities, the thermodynamic entropy (now the Noether
charge entropy, up to quantum corrections) is not simply a function of
area alone \cite{P09}. Although Penrose processes and the corresponding 
irreducible mass for black holes in these extended theories have not
been analyzed, we have shown that the Hilbert space description of black
hole evaporation implies that irreducible mass will be some function of
the Noether charge entropy. This suggests that if it is possible to
generalize Hawking's area theorem \cite{AreaThm} to these theories, then
some function of Noether charge entropy will instead play the role of
black hole surface area.

Our above analysis shows that the black hole tunneling probabilities
reduce to
\begin{equation}
\Gamma(\vec x|\vec X) ={\cal N}
e^{{\cal S}(\vec X-\vec x)-{\cal S}(\vec X)}; 
\label{simpleFinal}
\end{equation}
a result identical to the generic form of Eq.~(\ref{eq:formTunProb})
\cite{Haar}.
In a sense then, black holes are not ideal but ``real black bodies''
that satisfy conservation laws, result in a nonthermal spectrum and
preserve {\it thermodynamic\/} entropy. In contrast, 
%
treated as ideal black bodies, black hole evaporation would lead to
irreversible entropy production \cite{Zurek82}.

Our reasoning leading to Eq.~(\ref{simpleFinal}) holds for all
black hole and particle types (even in extended gravities) and is not
limited to one-dimensional WKB analyses which underlie all previous
quantum tunneling calculations. This supports our conjecture that
Eq.~(\ref{evapA}) provides a spacetime-free description of evaporation
across black hole horizons.

The physics deep inside the black hole is more elusive. 
Unfortunately, any analysis relying primarily on physics at
or across the horizon cannot shed any light on the question
of unitarity (which lies at the heart of the black hole information
paradox). If unitarity
holds globally, then Eq.~(\ref{evapA}) can be used to describe the
entire time-course of evaporation of a black hole and to learn how
the information is retrieved (see e.g., Ref.~\cite{BraunsteinA}).
Specifically, in a unitarily evaporating black hole, there should
exist some thermalization process, such that after what has been
dubbed the black hole's global thermalization (or scrambling) time, 
information that was encoded deep within the black hole can
reach or approach its surface where it may be selected for
evaporation as radiation.  Alternatively, if the interior of
the black hole is not unitary, some or all of this deeply encoded 
information may never reappear within the Hawking radiation. 

At this stage we might take a step back and ask the obvious question:
Does quantum information theory really bear any connection with the
subtle physics associated with black holes and their spacetime
geometry? After all we do not yet have a proper theory of quantum
gravity. However, whatever form such a theory may take, it should still
be possible to argue, either due to the Hamiltonian constraint of
describing an initially compact object with finite mass, or by appealing
to holographic bounds, that the dynamics of a black hole must be
effectively limited to a finite-dimensional Hilbert space. Moreover,
one can identify the most likely microscopic mechanism of black hole
evaporation as tunneling \cite{BraunsteinA,supmat}.
Formally, these imply that evaporation should look very much like
Eq.~(\ref{evapA}). Although finite, the dimensionalities of the Hilbert
space are immense and from standard
results in random unitary matrix theory and global conservation laws we
obtain a number of invariances. These invariances completely determine
the tunneling probabilities without needing to know the detailed
dynamics (i.e., the underlying Hamiltonian). This result puts forth
the Hilbert space description of black hole evaporation as a powerful
tool. Put even more strongly, one might interpret the analysis presented
here as a quantum gravity calculation without any detailed knowledge
of a theory of quantum gravity except the presumption of unitarity.

At a deeper level, the spacetime-free Hilbert space description and
random matrix calculus should apply to arbitrary event horizons, not
just those defining black holes (e.g., the Rindler horizon appears in
the infinite mass limit of the Schwarzschild geometry \cite{JP}).
In that case, Jacobson's work \cite{Jacobson,B} might suggest that the
gravitational structure of spacetime and presumably spacetime itself
along with related concepts could appear as emergent phenomena. If so,
the approach presented here may provide a promising beginning towards
achieving Verlinde's vision \cite{Verlinde}. However, to get even this
far required a subtle but crucial change in that vision. Rather than
emergence from a purely thermodynamic source, we should instead seek
that source in quantum information.

\vskip 0.05truein
\noindent
We thank S.\ Pirandola and N.\ Cohen for discussions.

\appendix*

\section{Preamble to the Hilbert space description}

In this section we provide background to the
introductory discussion of the manuscript.

\noindent
{\bf Causal separation implies a tensor product structure}:
Black holes are defined by their causal structure (their event horizons).
The event horizon specifies what is inaccessible from observation by an
external observer. In any quantum description of external observables
what is inaccessible must be traced out --- one necessarily has a tensor
product structure between the exterior and the remainder of Hilbert space
(the interior) ${\cal H}_{\text{ext}}\otimes {\cal H}_{\text{int}}$.

This observation is hardly new. It occurs automatically in field
theoretic descriptions. Indeed, such a tensor product structure was
explicitly utilized by Hawking \cite{Hawking76}. Further, it is exactly
what is seen in Rindler spacetime where the uniformly accelerated
observer has only access to signals on their side of the Rindler
event horizon --- tracing out the inaccessible degrees of freedom 
leaves a thermal state for the accelerated observer.

This use of the tensor product, to delineate what is outside and what
is not (at the Hilbert space level), in no way implies that the
spatial location of the event horizon cannot be fuzzy. These are
quite separate matters.

\noindent
{\bf The quantum mechanics of Hawking radiation}:
Whatever detailed field theoretic quantum gravity theory is ultimately 
developed, it is not unreasonable to expect that such a theory should 
allow for a description of black hole evaporation in terms of a microscopic 
(quantum mechanical) mechanism. As early as 1976, Hawking proposed
pair creation as this mechanism: Here, pair creation is conceived to 
occur outside the event horizon, with one of the pair falling into the
black hole (past the event horizon) and the other flying off as
Hawking radiation. The big
advantage of this mechanism is that it preserves the classical causal
structure of the black hole even at the quantum level --- Hawking's version
of a quantum black hole is of a perfectly `semi-permeable membrane' ---
anything can enter, nothing can leave; mass `escapes' because negative
energy is absorbed.

It was only very recently realized, however, that such a view is
completely at odds with the possibility of complete unitary
(quantum) evaporation of the black hole \cite{Nikolic}. Under
Hawking's mechanism each pair created will be pair-wise entangled
(entanglement between spin degrees of freedom; entanglement between
spatial degrees of freedom; indeed entanglement across all degrees
of freedom for the created pair). For each Hawking pair creation
event when one partner of the entangled pair passes the boundary
corresponding to the event horizon (as seen say by an infalling
observer) the rank of entanglement across that event horizon will
increase. Indeed, the structure of the tensor product provides a
natural framework for quantifying entanglement across the event horizon.

However, if the rank of entanglement across the event horizon is
increasing with each pair creation event then the Hilbert space
dimensionality of the black hole interior cannot vanish \cite{Nikolic}.
(We should note that the Hamiltonian constraint of describing an
initially compact object with a finite mass implies that the black
hole Hilbert space of any {\it dynamical\/} degrees of freedom must
be effectively finite dimensional.) This did not pose any obvious
problem for the {\it static\/} black hole spacetimes originally
considered by Hawking.  Rather, the difficulty is most glaring when
considering non-static black holes that can shrink and can eventually
vanish. Indeed, were Hawking's heuristic pair creation mechanism
correct the complete unitary evaporation of a black hole would be
utterly impossible \cite{Nikolic}. Here, we dub this catastrophic
inconsistency as `entanglement overload'.

For black holes to be able to eventually vanish, 
the original Hawking picture of a perfectly semi-permeable membrane must 
fail at the quantum level. In other words, entanglement overload very 
strongly points to the necessary breakdown of the classical causal 
structure of a black hole. This statement already points to the
likely solution.

\noindent
{\bf Evaporation as tunneling}:
The most straightforward way to evade entanglement overload is
for Hilbert space within the black hole to `leak away' --- quantum 
mechanically we would call such a mechanism tunneling \cite{Braunstein}.
Indeed, for over a decade now, such tunneling, out and across the event 
horizon, has been used as a powerful way of computing black hole
evaporation rates including the effects of backreaction.

We suggest that the evaporation across event horizons operates
by Hilbert space subsystems from the black hole interior moving to
the exterior. The equation
\begin{equation}
|i\rangle_{\text{int}} \rightarrow (U|i\rangle)_{BR},
\label{evap}
\end{equation}
[Eq.~(3) of the manuscript]
provides the simplest mechanism for this to occur: Subsystems
are dynamically selected (by some unitary $U$) and reassigned as
radiation in an enlarged exterior Hilbert space.

\noindent
{\bf Spacetime free conjecture:}
This brings us to the key conjecture of the manuscript: that
Eq.~(\ref{evap}) above (all equation numbers herein refer to
Supplementary Material equations unless explicitly referring back
to the manuscript) accurately describes the evaporation across black
hole event horizons.

Our manuscript primarily investigates the consequences of Eq.~(\ref{evap})
applied specifically to event horizons of black holes. Now the consensus
appears to be that the physics of event horizons (cosmological, black
hole, or those due to acceleration) is universal. In fact, it is precisely 
because of this generality that one should not expect Eq.~(\ref{evap}) 
to bear the signatures of the detailed physics of black holes. Rather
we then go on to impose the details of that physics onto this equation.

\noindent
{\bf Testing this conjecture:}
The manuscript is devoted to exploring the implications of
Eq.~(\ref{evap}) for the evaporation rates of black holes, thus
providing a test of its predictive power. To achieve
this, the key pieces of physics about black holes
we rely on are the no-hair theorem and the existence of Penrose
processes. We assume that any quantum representation of a black hole
must have a direct correspondence to its classical counterpart where
these properties hold true. Therefore, when we wish to apply the
very general Hilbert space description of quantum tunneling across
event horizons in Eq.~(\ref{evap}) we need to impose
conditions consistent with these classical properties of a black hole. 

It is our contention that the key technical content of the manuscript
[involving its Theorems 1 through 4 and leading to its Eq(14)] provides
strong evidence in support of the conjecture that Eq.~(\ref{evap})
describes the evaporation across black hole event horizons.
Importantly, the generality of this equation suggests that evidence 
which supports the validity of Eq.~(\ref{evap}) for black holes 
likely implies its more universal validity as a description of
evaporation across arbitrary event horizons.

\section{Technical proofs and minor notes}

\noindent
{\bf Proof of Theorem 1:}
We observe that it is trivial to verify that any function
$\Gamma(\varepsilon|M)$ of
the form $e^{f(M-\varepsilon)-f(M)+h(\varepsilon)}$
satisfies 
$\Gamma(\varepsilon_1|M)\, \Gamma(\varepsilon_2|M-\varepsilon_1)
=\Gamma(\varepsilon_2|M)\, \Gamma(\varepsilon_1|M-\varepsilon_2)$.
To prove this is the general solution set
$\gamma(\varepsilon,M)=\ln{\Gamma(\varepsilon|M)}$. Then
$\gamma$ satisfies an additive equation
\begin{equation}
\label{eq:ampScalarFunc2}
\gamma(\varepsilon_1,M-\varepsilon_2)+\gamma(\varepsilon_2,M)
= \gamma(\varepsilon_2,M-\varepsilon_1)+\gamma(\varepsilon_1,M).
\end{equation}
Taking the partial derivative of this equation with-respect-to
$\varepsilon_2$ and then setting $\varepsilon_1=\varepsilon$ and
$\varepsilon_2=0$ yields
\begin{equation}
\label{eq:basic_pdiff} 
\gamma_2(\varepsilon,M)=\gamma_1(0,M)-\gamma_1(0,M-\varepsilon),
\end{equation}
where 
$\gamma_1(\varepsilon,M)
\equiv\partial\gamma(\varepsilon,M)/\partial \varepsilon$ and
$\gamma_2(\varepsilon,M)
\equiv\partial\gamma(\varepsilon,M)/\partial M$.
A general solution to this equation is given by 
\begin{equation}
\gamma(\varepsilon,M)= 
\!\!\int_{M-\varepsilon}^\infty \!\gamma_1(0,M')\,\diffD M' -
\int_M^\infty\!\gamma_1(0,M')\,\diffD M' +h(\varepsilon),
\end{equation}
where $h(\varepsilon)$ is an arbitrary function. Now setting 
\begin{equation}
f(M)=\int_M^\infty\gamma_1(0,M')\,\diffD M',
\end{equation}
we have $\gamma(\varepsilon,M)=f(M-\varepsilon)-f(M)+h(\varepsilon)$. 
{$~$}\hfill \rule{2mm}{2mm}
\vskip 0.1truein

\noindent
{\bf Proof of Theorem 2:}
The case of energy and charge which are scalar observables is obvious. 
We have to consider angular momentum only. As is well-known angular 
momentum operators generate the Lie algebra $su(2)$. The
finite-dimensional representations of  this algebra are {\em completely 
reducible}, that is, the state space can be decomposed into a direct 
sum of {\em irreducible} representations. Moreover, since the black 
hole is to be in a definite angular momentum state each of the summands
must have the same $J^2$ eigenvalue. It therefore suffices to focus 
on any one irreducible summand. We will freely use the standard 
properties of irreducible representations. Since the post-evaporation 
state of a black hole must be a spin-coherent state and the only 
orthogonal set of spin-coherent states are of the form
$\{R(\theta,\phi)\ket{j,j}, R(\theta,\phi)\ket{j,-j}\}$ the general 
state of the evaporated particle and black hole is given by
\begin{eqnarray}
\Phi &=& \alpha'\otimes R(\theta,\phi)\ket{j,j}+ \beta'\otimes
R(\theta,\phi)\ket{j,-j} \\
&=& R(\theta,\phi)\otimes R(\theta,\phi)
(\alpha\otimes \ket{j,j}+ \beta\otimes \ket{j,-j}), \nonumber
\end{eqnarray}
where $R(\theta,\phi)\alpha=\alpha'\text{ and }R(\theta,\phi)\beta=\beta'$
denote (unnormalized) states of the evaporated particle. Let the
operator representing $J^2$ on the product space be denoted
$J_{\text{tot}}^2$. Then the condition that $\Phi$ be an eigenstate
of $J^2\otimes \openone$, $\openone\otimes J^2$ and $J_{\text{tot}}^2$
implies it must be an eigenstate of the operator 
\begin{eqnarray}
\tilde{J}&\equiv& J_{\text{tot}}^2-J^2\otimes \openone
-\openone\otimes J^2 \nonumber \\
&=& J_+\otimes J_-+ J_-\otimes J_+ + 2J_z\otimes J_z.
\end{eqnarray}
As $\tilde{J}$ is invariant under $R(\theta,\phi)\otimes R(\theta,\phi)$
this implies that 
\begin{eqnarray}
\label{eq:j_condition}
&&J_+\alpha\otimes J_-\ket{j,j}+ J_-\beta\otimes J_+ \ket{j,-j}\nonumber \\
&&+2j(J_z\alpha\otimes \ket{j,j} -J_z\beta\otimes \ket{j,-j})\nonumber \\
&=& x(\alpha\otimes \ket{j,j} + \beta\otimes \ket{j,-j})
\end{eqnarray}
where $x$ is a real number.

First, suppose that $j>1$. Then the vectors $\ket{j,j}$, $\ket{j,-j}$,
$J_-\ket{j,j}$ and $J_+\ket{j,j}$ are mutually orthogonal and the
above equation can be satisfied if and only if either
$\beta=0$ and $J_+\alpha=0$ or $\alpha=0$ and $J_-\beta=0$. We conclude
that in this case the only allowed forms of
$\Phi$ are (up to a global rotation) $\ket{j_p,j_p}\otimes \ket{j,j}$ and
$\ket{j_p,-j_p}\otimes \ket{j,-j}$ where $\ket{j_p,j_p}$
($\ket{j_p,-j_p}$) is the highest (lowest) eigenvector in the particle's
angular momentum space. Clearly these  states are always $J^2$
eigenstates for any value of $J$. We call such states for $\Phi$ standard. 
To conserve $J_{\text{tot},\hat{n}}$ the state of the mother black hole
must be $R(\theta,\phi)\ket{j+j_p,\pm(j+j_p)}$ respectively.

Next suppose $j=1$. It follows from Eq.~\eqref{eq:j_condition}
that besides the standard states the state (up to a global rotation)
\begin{equation}
\frac{1}{\sqrt{2}}
(\ket{j_p,-1}\otimes\ket{1,1}-\ket{j_p,1}\otimes\ket{1,-1}),
\end{equation}
is also an eigenstate of $J_{\text{tot}}^2$ with total angular momentum
of the mother black hole $j'=j_p$. As this is a $J_{\text{tot},\hat{z}}$
eigenstate with zero eigenvalue no orientation can conserve
$J_{\text{tot},\hat{n}}$ of the original black hole. We therefore rule
this class of states out.

Next suppose $j=\frac{1}{2}$. In addition to the standard states there
are other possibilities. The product space decomposes into two
irreducible representations corresponding to total angular momentum
$j'=j_p\pm \frac{1}{2}$. They are generated respectively by highest
weight vectors (up to a global rotation)
\begin{equation}
\ket{j_p,j_p} \otimes
\ket{{\textstyle \frac{1}{2}},{\textstyle \frac{1}{2}}},
\end{equation}
for $x=j_p$ corresponding to $j'=j_p+\frac{1}{2}$ and
\begin{equation}
\frac{1}{\sqrt{2j_p+1}}\bigl( \sqrt{2j_p}\ket{j_p,j_p}
\otimes\ket{{\textstyle \frac{1}{2}},-{\textstyle \frac{1}{2}}} 
- \ket{j_p,j_p-1} \otimes
\ket{{\textstyle \frac{1}{2}},{\textstyle \frac{1}{2}}}\bigr),
\end{equation}
for $x=-1-j_p$ corresponding to $j'=j_p-\frac{1}{2}$. Starting with
either of these we can generate the other ${J}_{\hat z}$ eigenvectors
(in this globally rotated basis) by successive applications of the
${J}_-$ operator. However, considering conservation of
${J}_{\text{tot},\hat{n}}$ disallows any of these extra eigenvectors.
Therefore, when quantized along the $\hat n$ axis the mother black hole
had the state $\ket{j_p+\frac{1}{2},j_p+\frac{1}{2}}$ and
$\ket{j_p-\frac{1}{2},j_p-\frac{1}{2}}$ respectively (with the exception
of the case $j_p=\frac{1}{2}$  for the latter mother black hole state
with $j'=0$ where the orientation of the quantization axis is arbitrary).

This leaves only $j=0$ which is trivial. It is now easy to
check that in every case allowed by global conservation laws
the statement of the theorem holds true.
{$~$}\hfill \rule{2mm}{2mm}
\vskip 0.1truein

\noindent
{\bf Proof of Theorem 3:}
Let $\gamma(\vec x,\vec X)=\ln \Gamma(\vec x|\vec X)$. Then $\gamma$ satisfies
\begin{equation}
\label{eq:basicFunc2_multi}
\gamma(\vec x,\vec X)+\gamma(\vec x',\vec X - \vec x)
=\gamma(\vec x',\vec X)+\gamma(\vec x,\vec X - \vec x')
\end{equation}
Taking the partial derivative with-respect-to $x'_i$ and then setting
$x'_i=0$ yields
\begin{equation}
\label{eq:basic_pdiff2}
\frac{\partial \gamma(\vec x,\vec X)}{\partial X_i}
=\frac{\partial \gamma}{\partial x_i}\Bigr|_{(\vec 0,\vec X)}
-\frac{\partial \gamma}{\partial x_i}\Bigr|_{(\vec 0,\vec X-\vec x)}.
\end{equation}
As in Theorem 1, the solution to the above partial differential equation
for $i=n$ can be inferred from Eq.~\eqref{eq:basic_pdiff} above by
treating all variables except the last ``conjugate'' pair
$(x_n,X_n)$ as constants, so
\begin{equation}
\gamma(\vec x,\vec X)=f_n(\vec X-\vec x) - f_n(\vec X) + h_n(\vec x, \hat X)
\end{equation}
where $\hat X=\{X_1,\ldots,X_{n-1}\}$ without dependence on $X_n$;
the function $h_n(\vec x, \hat X)$ is otherwise arbitrary.
Now, substituting this into the functional equation
\eqref{eq:basicFunc2_multi} for $\gamma$ and noting that 
\begin{equation}
f_n(\vec X-\vec x) - f_n(\vec X)
\end{equation}
is a already a solution of it, we see that $h_n$ satisfies the equations
\begin{equation}
\label{eq:func_reduced}
\frac{\partial h_n(\vec x,\hat X)}{\partial X_i}
=\frac{\partial h_n}{\partial x_i}\Bigr|_{(\vec 0,\hat X)}
-\frac{\partial h_n}{\partial x_i}\Bigr|_{(\vec 0,\hat X-\hat x)}
\end{equation}

Now consider the function $\partial h_n(\vec x, \hat X)/\partial x_n$
we have 
\begin{eqnarray}
&&\frac{\partial}{\partial X_i}
\Bigl(\frac{\partial h_n(\vec x,\hat X)}{\partial x_n}\Bigr)
=\frac{\partial}{\partial x_n}
\Bigl(\frac{\partial h_n(\vec x,\hat X)}{\partial X_i}\Bigr)\nonumber \\
&=&\frac{\partial}{\partial x_n}
\Bigl(\frac{\partial h_n}{\partial x_i}\Bigr|_{(\vec 0,\hat X)}
-\frac{\partial h_n}{\partial x_i}
\Bigr|_{(\vec 0,\hat X-\hat x)}\Bigr)\equiv 0,
\end{eqnarray}
since $\hat x$ has no dependence on $x_n$.
Hence the function $\partial{h_n(\vec x, \hat X)}/\partial{x_n}$
can have no dependence on any $X_i$.
Consequently the function $h_n(\vec x,\hat X)$ must have the form 
\begin{equation}
\label{eq:funcFormRed}
h_n(\vec x,\hat X)= u_n(\vec x) +\gamma_{n-1}(\hat x, \hat X).
\end{equation}
The function $\gamma_{n-1}$ satisfies the functional equation
\eqref{eq:basicFunc2_multi} with $n-1$ pairs of conjugate variables.
Hence
\begin{equation}
\gamma(\vec x,\vec X)= f_n(\vec X-\vec x)-f_n(\vec X) + u_n(\vec x)
+ \gamma_{n-1} (\hat x, \hat X).
\end{equation}
Using this argument recursively and absorbing the different
functions together, we conclude that 
\begin{equation}
\gamma(\vec x,\vec X)=f(\vec X-\vec x)-f(\vec X) + h(\vec x).
\end{equation}
{$~$}\hfill \rule{2mm}{2mm}

\noindent
{\bf Note:}
From Theorem~{4}, permutation symmetry yields
\begin{equation}
\label{tunProbApp}
\Gamma(\vec x|\vec X)= e^{f(\vec X-\vec x)-f(\vec X)+h(\vec x)}.
\end{equation}
For infinitessimal $\vec x$ backreaction should be negligible and we
should recover the Hawking thermal spectrum, i.e.,
\begin{equation}
\Gamma(\vec x|\vec X)\simeq
e^{-\vec \nabla f(\vec X)\cdot \vec x+h(\vec 0)} 
\equiv N e^{-\vec \nabla {\cal S}(\vec X)\cdot \vec x},~~~
\forall \vec X.
\end{equation}
Here ${\cal S}(\vec X)$ is the thermodynamic entropy of the black hole,
$N$ is a normalization constant and without loss of generality we have
absorbed any linear part of $h$ into $f$. Solving
$\vec \nabla f(\vec X)=\vec \nabla {\cal S}(\vec X)$
yields $f(\vec X)={\cal S}(\vec X)$ since $f(\vec 0)$ may be chosen
arbitrarily. Note that the reasoning provided in the manuscript does
{\it not\/} rely on this argument {\it nor\/} on consistency with the
Hawking thermal spectrum.
\vskip 0.1truein

\noindent
{\bf Proof of Theorem~4:} Let $\vec X\in \Sigma^o-K$, then by
definition there is some component $X_i$ of $\vec X$ such that
$\partial {\cal I}/\partial X_i\neq 0$ at $\vec X$. Without loss of
generality we may take $i=n$. Then there is some neighborhood $O$ of
$\vec X$ such that $\partial {\cal I}/\partial X_n\neq 0$ at every point
in $O$. Consider the continuously differentiable map $F: O\rightarrow O$
\begin{equation}
F(\vec{X}) = (X_1,\ldots,X_{n-1}, {\cal I}(\vec{X})).
\end{equation}
The Jacobian of $F$ is simply $|\partial {\cal I}/\partial X_n|$ 
and does not vanish anywhere in $O$. From the inverse function theorem
then there is a neighborhood $\tilde O\subset O$ such that $F$ is
{\it invertible\/} in $\tilde O$. Thus any $\vec X\in \tilde O$ can be
written in the new coordinate system as
$\vec X = (X_1,\ldots, X_{n-1}, {\cal I}(\vec X))$. Let $\theta$ be
the corresponding function that represents $\Theta$ in the new coordinates.
Then for $\vec X_1, \vec X_2\in \tilde O$ and
$\vec X_1', \vec X_2'\in \tilde O'$ the hypothesis in Eq.~(10) [of
the manuscript] is equivalent to
$\theta(X_{1,1},\ldots, X_{1,n-1}, {\cal I}(\vec X_1),
X_{1,1}',\ldots, X_{1,n-1}', {\cal I}(\vec X_1'))
=\theta(X_{2,1},\ldots, X_{2,n-1}, {\cal I}(\vec X_2),
X_{2,1}',\ldots, X_{2,n-1}', {\cal I}(\vec X_2'))$
But this is precisely the statement that $\theta$ is independent of
the first $n-1$ coordinates in each argument.
Hence $\Theta(\vec X,\vec X') = \theta({\cal I}(\vec X),{\cal I}(\vec X'))$
in $\tilde O\times \tilde O'$. This must be true for every pair of
points in $\Sigma^o-K$. Note that although for another pair of points
say $\vec Y,\vec Y'\in \Sigma^o-K$ the new $\theta_Y$ may be a different
function, $\theta$ and $\theta_Y$ must match in any common domain
since $\Theta$ is globally defined. Hence there is a continuously
differentiable function $\theta$ such that the assertion of the
theorem holds for any pair of arguments in $\Sigma^o-K$. Since the latter
is a dense subset of $\Sigma$, $\theta$ can be uniquely extended to the
whole of $\Sigma\times \Sigma$ by continuity.
{$~$}\hfill \rule{2mm}{2mm}
\vskip 0.1truein

\noindent
{\bf Note:}
The irreducible mass of a black hole with no-hair triple
$\vec X=(M,Q,J)$ in General Relativity is 
\begin{equation}
{\cal I}=\frac{1}{2}\,
\Bigl(2M^2-Q^2 +2M\sqrt{M^2-Q^2-a^2}\Bigr)^{\!{1}/{2}},
\end{equation}
where $a\equiv J/M$. It is straightforward to check that this function
satisfies the condition in Theorem~4 that
$\{\vec X:|\vec \nabla {\cal I}(\vec X)|=0\}$ is nowhere dense.
\hfill
\vskip 0.1truein

\noindent
{\bf Note:}
It has been noted in the literature \cite{Zhang,P04,Arzano}
that Eq.~(1) [of the manuscript] for the Schwarzschild case naively
satisfies the relation \cite{fnn}
\begin{equation}
\Gamma(\varepsilon_1|M)\Gamma(\varepsilon_2|M-\varepsilon_1)
=\Gamma(\varepsilon_1+\varepsilon_2|M).
\label{new_fe}
\end{equation}
by symmetry of $\varepsilon_1+\varepsilon_2$ it is trivial to use
Theorem~{1} from our manuscript to write down
the general solution to Eq.~(\ref{new_fe}) as
\begin{equation}
\Gamma(\varepsilon|M)
\equiv e^{f(M-\varepsilon)-f(M)},
\end{equation}
for some function $f$.

\noindent
{\bf Note:}
Consider one form of the well-known Cauchy functional equation \cite{Aczel}
\begin{equation}
G(a)G(b) = G(a+b).
\label{Ceqn}
\end{equation}
Its unique solution is the exponential family of functions.

Naively, Eq.~(\ref{new_fe}) is apparently a natural generalization to the
Cauchy equation~(\ref{Ceqn}) when incorporating conservation laws.
However, as already noted \cite{fnn} its interpretation is problematic.
By contrast, the functional equation
\begin{eqnarray}
\Gamma(\vec x|\vec X)\,\Gamma(\vec x'|\vec X - \vec x)
&=&\Gamma(\vec x'|\vec X)\,\Gamma(\vec x|\vec X - \vec x'),
\end{eqnarray}
[Eq.~(7) of the manuscript] provides a truly non-trivial generalization to
the Cauchy functional equation in the presence of conservation laws. Its
interpretation is clear as a permutation symmetry (see manuscript)
and further it includes Eq.~(\ref{new_fe}) as a special case.


\begin{thebibliography}{99}
\bibitem{Haar} Although Eq.~(\ref{eq:formTunProb}) has been
obtained to leading orders in $\hbar$ in WKB calculations,
Eq.~(\ref{simpleFinal}) is exact to all orders (if the Haar symmetry
is exact). However, as with all tunneling calculations it ignores
gray-body factors \cite{Bek94}.
\bibitem[S7]{fnn} In fact, this is possible only if the normalization
is excluded or absorbed into the measure. More worrying, perhaps, is
the interpretation of this relation. Why for example does it not
include a sum over all ways of partitioning the total energy
$\varepsilon_1+\varepsilon_2$ into two (or more) sub-components?
\end{thebibliography}
\end{document}